\documentclass[pra,aps,reprint,showpacs,amsmath,superscriptaddress]{revtex4-1}

\usepackage{bm}
\usepackage{graphicx}
\usepackage{dsfont}

\usepackage{mathrsfs}

\usepackage{framed}
\usepackage{multirow}
\usepackage[all]{xy}

\usepackage{framed}
\usepackage{comment}
\usepackage{here}                    
\usepackage{latexsym}		     
\usepackage{amsmath}     
\usepackage{amsfonts}  
\usepackage{amssymb}
\usepackage{amsthm}
\usepackage{dcolumn}
\usepackage{color}
\usepackage{mathrsfs}
\usepackage{dsfont}

\renewcommand{\phi}{\varphi}

\newcommand{\be}{\begin{equation}}
\newcommand{\ee}{\end{equation}}
\newcommand{\bea}{\begin{eqnarray}}
\newcommand{\eea}{\end{eqnarray}}

\renewcommand{\phi}{\varphi}

\usepackage[colorlinks,citecolor=blue,linkcolor=blue,urlcolor=blue]{hyperref}

\begin{document}

\title{Remote quantum-safe  authentication of entities with physical unclonable functions}

\author{Georgios M. Nikolopoulos}
\email{nikolg@iesl.forth.gr}

\affiliation{Institute of Electronic Structure and Laser, Foundation for Research and Technology-Hellas, GR-70013 Heraklion, Greece}

\date{\today}

\begin{abstract} 
Physical unclonable functions have been shown a useful resource of randomness for implementing various cryptographic tasks including entity authentication. All of the related entity authentication protocols that have been discussed in the literature so far, either they are vulnerable to an emulation attack, or they are limited to short distances. 
Hence, quantum-safe remote entity authentication over large distances remains an open 
question. In the first part of this work we discuss the requirements that an entity authentication protocol has to offer 
in order to be useful for remote entity authentication in practice. 
Subsequently, we propose a protocol, which can operate over 
large distances, and offers security against both classical and quantum adversaries.  
The proposed protocol relies on standard techniques, it is fully compatible with  the infrastructure of existing and future photonic networks, and it can operate in parallel with other quantum protocols, including QKD protocols.
\end{abstract}


\maketitle


\section{Introduction}

Entity authentication is one of the main pillars of our digital world, which is 
widely employed to control access of users to physical or virtual resources 
 \cite{book1,book2}. 
The typical scenario involves a human user (claimant) who has to provide 
in real time evidence, of her identity to another human or non-human 
entity (verifier), 
in order to obtain access e.g., to a laboratory or to a bank account. 

Entity authentication can be provided by means of various 
techniques, some of which are not inherently cryptographic. 
Among the cryptographic techniques,  dynamic schemes with a 
challenge-response mechanism are of particular interest, because 
they offer high level of security for most everyday tasks \cite{book1,book2}. This is, for instance, the case of smart cards (tokens) that  are widely used e.g., in transactions through ATMs, as well as in e-commerce. To a large extent, their security relies on a short (typically four- to eight-digit) PIN, which is connected to the card and it is known to the legitimate owner of the card, as well as an independent long numerical  secret key, which is stored on the card, and the verifier has a matching counterpart of it. The PIN may also be stored on the token, and it provides an additional level of security in case the token is lost or stolen. A common technique is for the PIN to serve as a means for the verification of the  user to the token, and subsequently the token (representing the user) authenticates itself to the system by 
means of the additional secret key stored on it (e.g., the token may be asked to encrypt a randomly chosen challenge with its key). Hence, we have a two-stage authentication, which requires the user to remember the short PIN, and to possess the token where the longer secret key is stored. There are cases where both of these stages take place (e.g., transactions through ATMs), and cases where the authentication of the user to the card suffices (e.g., e-commerce). In either case, the secrecy of the long numerical secret key is of vital importance, and 
there are various physical invasive and non-invasive attacks as well as software  attacks,  which aim at its extraction  from the card  \cite{PUFtutorial2014,holcomb2014pufs,Lim-etal:2005,handbook3}. 

Physical unclonable functions (PUFs) have been proposed as a means for developing entity 
authentication protocols (EAPs) \cite{Pappu02,PappuPhD}, 
which do not require the storage of a secret numerical key, and thus they are resistant to the aforementioned attacks. 
Typically, a PUF relies on a physical token with internal randomness, which is  introduced explicitly or 
implicitly during its fabrication, and it is considered to be technologically hard to clone (hence the 
term physical unclonable). The operation of PUF-based EAPs relies on a challenge-response 
mechanism, 
where the verifier accepts or rejects the identity of the claimant based on the responses of the token 
to one or more randomly chosen challenges. A list of challenge-response pairs is generated 
by the manufacturer, before the token is given to a user, and it is stored in a system database 
where the verifier has access over an authenticated classical channel.  

The nature of the physical token essentially determines the nature of the challenges and the responses, as well as the operation of the PUF. 
One can find a broad range of PUFs in the literature \cite{PUFtaxonomy:2019,Gao20,Chow21}, and  
electronic PUFs is the most well studied class, 
mainly because of their compatibility with existing technology and hardware. However, they are susceptible to various types of modelling and side-channel attacks  \cite{cryptoeprint:2019:566,khalafalla2019pufs,ganji2019pufmeter,delvaux2019machine,Chow21},  
and current research is focused on the development of new schemes, which offer 
provable security against such attacks (e.g., see \cite{gunlu} and references therein). 
On the contrary, optical PUFs are not fully compatible with existing technology, but they offer many advantages relative to electronic PUFs, including low cost, high complexity, and  security against modeling attacks \cite{opticalPUF2013,Chow21}. Typically, their operation relies on the 
response of a disordered optical multiple-scattering medium (token), when probed by light with randomly chosen parameters. In general, one can distinguish between two major classes of optical PUFs. In optical PUFs with classical readout, different numerical challenges are encoded on the parameters of  laser light that is scattered by the token \cite{Pappu02,PappuPhD,Horstmayer13,Mes18}. Typically, such encoding may involve the wavelength, the wavefront, the point and angle of incidence, or a combination thereof.  In the second major class of optical PUFs, the token is interrogated 
by quantum states (quantum readout), and different numerical challenges are encoded on non-orthogonal quantum states of light \cite{Skoric12,Goorden:14,NikDiaSciRep17,Nik18,FlaNikAlbFis18,Nikolopoulos:19}.  

To the best of our knowledge, all of the known EAPs with classical readout of 
optical-PUFs are susceptible to an emulation attack \cite{Skoric12,Goorden:14}. 
On the other hand, as explained below in more detail, 
all of the known EAPs that rely on optical PUFs and offer 
security against  an emulation attack \cite{NikDiaSciRep17,Nik18,FlaNikAlbFis18}, assume a tamper-resistant verification setup, while they are 
limited to short distances (typically $<10$ km). 
Moreover, none of the above EAPs (with classical \cite{Pappu02,PappuPhD,Horstmayer13,Mes18} or quantum readout \cite{Skoric12,Goorden:14,NikDiaSciRep17,Nik18,FlaNikAlbFis18,Nikolopoulos:19})  takes into account the possibility of the token to be 
stolen or lost, and thus they are not secure in a scenario where the token is 
possessed by the  adversary. 
In the present work our aim is to propose a protocol which addresses all of these issues, 
and it offers quantum-secure entity authentication over  large distances ($\gtrsim 100$ km). 
The proposed protocol does not require the storage of any secret key, apart from the short PIN which accompanies the token and it has to be memorized by the holder. Moreover, the protocol can be implemented with current technology, and it is fully compatible with existing and forthcoming 
quantum key-distribution (QKD) infrastructure. 



\section{Materials and Methods}
\label{secII}

For the sake of completeness, in this section we summarize briefly the main optical-PUF-based EAPs that have been discussed in the literature so far, focusing on their vulnerabilities in a remote authentication scenario, which is the main subject of the present work. 

The general setting of an EAP involves two parties: the prover or claimant 
(Alice) and the verifier (Bob).  
Alice (A) presents evidence about her identity to Bob (B), and the main task of Bob is to confirm the claimed identity. 
Typically, an EAP may rely on something that the claimant knows (e.g., a password), 
something that the claimant possesses (e.g., a key), or on a combination of the two. To avoid any misunderstandings, throughout this work Alice is considered to be honest, whereas there is a third party (Eve), who intends to 
impersonate Alice to the verifier. 

Depending on whether the optical challenges are formed by encoding numerical challenges on classical or quantum states of light, one can distinguish between optical PUFs with classical  \cite{Pappu02,PappuPhD,Horstmayer13,Mes18} and quantum readout \cite{Skoric12,Goorden:14,NikDiaSciRep17,Nik18,FlaNikAlbFis18}. 
In either case, all of the related EAPs that have been 
proposed in the literature,  rely on a challenge-response mechanism and the existence of a challenge-response database, which is  generated by the manufacturer only once, before the token is given to the legitimate user. 
The database characterizes fully the response of 
the token with respect to all the  possible challenges that may be chosen by the system, and thus the verifier may accept or reject the token based solely on its response to a finite number of randomly chosen challenges. 
In a single authentication session, the token is inserted in a verification set-up, where the verifier has remote access, and it is interrogated by a number of randomly chosen optical challenges. The corresponding responses are returned to the verifier, and they are compared to the expected ones. The verifier accepts the  identity of the claimant if the recorded responses are compatible with the expected ones, and rejects it otherwise. 

To the best of our knowledge, besides the unclonability of the token,  all of the known optical EAPs  assume that the token is in possession of the legitimate user \cite{Pappu02,PappuPhD,Horstmayer13,Mes18,Goorden:14,NikDiaSciRep17,Nik18}. 
However, there is nothing in these protocols that binds the token to the 
legitimate owner, which means that whoever has the token can impersonate successfully 
the legitimate owner to the system \cite{book1}. 
In other words, none of the existing schemes is secure when the token is lost or stolen. 
In analogy to conventional smart cards, one way to prevent the use of the token by an unauthorized user is to introduce a PIN 
which is connected somehow to the token, and it is memorized by the legitimate owner of the token. Moreover, to facilitate remote entity authentication, it is desirable for the PIN verification to be possible without any  access to 
a central database. In conventional smart cards this is achieved by storing the PIN on the  chip of the card. However, the generalization of this approach to all-optical EAPs is not straightforward, 
and one has to design judiciously the protocol.

Communication between the claimant and the verifier  is inevitable for any EAP, irrespective of whether this is performed by means of classical or quantum resources. We are interested in 
EAPs which remain secure even when an adversary monitors this communication. One way to satisfy this requirement is to ask for Alice and Bob to be connected via  trusted communication line, which in turn imposes additional requirements (e.g., additional secret keys).  The assumption of trusted communication line, safe from monitoring, may be reasonable for a local authentication scenario, where Alice and Bob are in the same building, but it is not suitable for remote entity authentication, where the information exchanged between them may travel  hundreds of kilometers over open communication lines. In this case, the security of entity authentication requires 
guarding against potential adversaries who monitor the communications, and the development of secure protocols becomes rather challenging.

\begin{figure*}
\centering\includegraphics[scale=0.6]{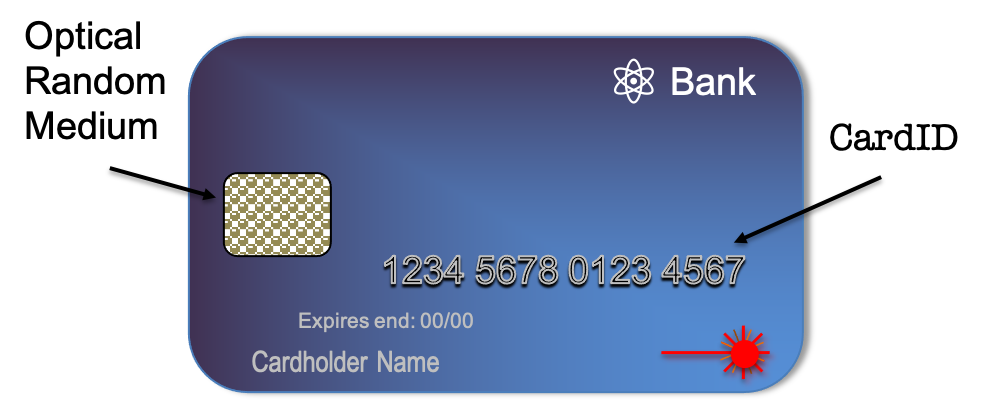}
\caption{Schematic representation of an optical token. 
The token resembles a conventional smart card, where the 
chip is replaced by a disordered optical  multiple-scattering medium. 
}
\label{card:fig}
\end{figure*}

Schemes with classical readout \cite{Pappu02,PappuPhD,Mes18}, have an advantage over schemes with quantum readout, in the sense that they are not susceptible to inevitable losses and noise associated with the transmission of quantum states over large distances. As a result, they can operate over arbitrarily large distances, but they are vulnerable to emulation and replay attacks \cite{book1,Goorden:14}, if an adversary (Eve) obtains undetected access to the database of challenge-response pairs, or if she monitors the classical communication between Alice and Bob.  

To prevent such  attacks, various authors have proposed EAPs in which different numerical challenges are mapped onto non-orthogonal quantum states of light, and the authentication of the token involves one or more such states chosen at random and independently \cite{Skoric12,Goorden:14,NikDiaSciRep17,Nik18,FlaNikAlbFis18}. In this case, in order for Eve 
 to impersonate successfully Alice (without  access to her token), she has to identify successfully each one of the transmitted states, and to send the right sequence of responses to the system.  Fundamental laws of quantum physics limit the amount of information that can be extracted from each state, and it is inevitable for Eve to deduce the wrong  challenge for some of them. In these cases she will send the wrong responses to the verifier, and her intervention will be revealed. 
These ideas are exploited by the protocols in Refs.  \cite{Skoric12,Goorden:14,NikDiaSciRep17,Nik18,FlaNikAlbFis18}, and the security of the protocols has been investigated against various types of intercept-resend attacks, where the adversary has obtained access to the database of challenge-response pairs, and the verification set-up is {\em tamper resistant}. 
The latter assumption implies that all of the components of the verification set-up are controlled by the verifier, and the adversary has access to the optical challenge only immediately  before it impinges on the token. 

Unfortunately, for practical reasons, all of these schemes are limited to short distances ($<10$km). This is well below the distances that can be covered nowadays by standard QKD protocols, which employ single-mode fibers (SMFs) \cite{QKDreview1,QKDreview2}. More precisely, extension of the protocol of Goorden {\em et al.} \cite{Goorden:14} to large distances requires the use of multimode fibers, which allow for the  transmission of the modified wavefront from the verifier to the claimant and backwards. Inevitable time-dependent variations of the multimode fiber during the transmission of the optical signals will result in cross-mode coupling 
(e.g., see \cite{MMFlosses} and references therein), thereby limiting the distances over which secure entity authentication can be achieved.  
Moreover, integration of the authentication scheme of Goorden {\em et al.} 
in forthcoming quantum communications infrastructures, is complicated considerably by the necessity for reliable low-loss interfaces between multi-mode fibers and SMFs. 
On the other hand, although the scheme of  Refs. \cite{NikDiaSciRep17,Nik18,FlaNikAlbFis18} 
is fully compatible with existing QKD infrastructure, it relies on a  Mach-Zehnder interferometric setup. As a result, it can operate reliably only if  the relative length of the arms in the interferometer does not change by more than a fraction of a wavelength \cite{QKDreview1}. This is also possible for short distances ($<10$km), but it is getting harder when the distance between the verifier and the claimant increases, because of inevitable environmental variations. 

In the following sections we propose an EAP, which addresses all of the above issues and it is suitable for remote 
authentication over arbitrary distances. At the core of our scheme there is a token ${\cal T}$, which is given to a user 
and it is used for her 
identification by the verifier. The token resembles a  standard smart card in conventional EAPs, with a random multiple-scattering  optical medium in  place of the chip (see Fig. \ref{card:fig}). 
The faithful cloning of the token typically requires the exact positioning (on a nanometer scale) of millions of scatterers with the exact
 size and shape, which is considered to be a formidable challenge not only for current, but for future technologies 
as well. Hence, the internal disorder on the one hand renders the cloning of the token 
a formidable challenge, while on the other hand may serve as 
a physical source of randomness for the development of cryptographic primitives. 

\begin{figure*}
\centering\includegraphics[scale=0.6]{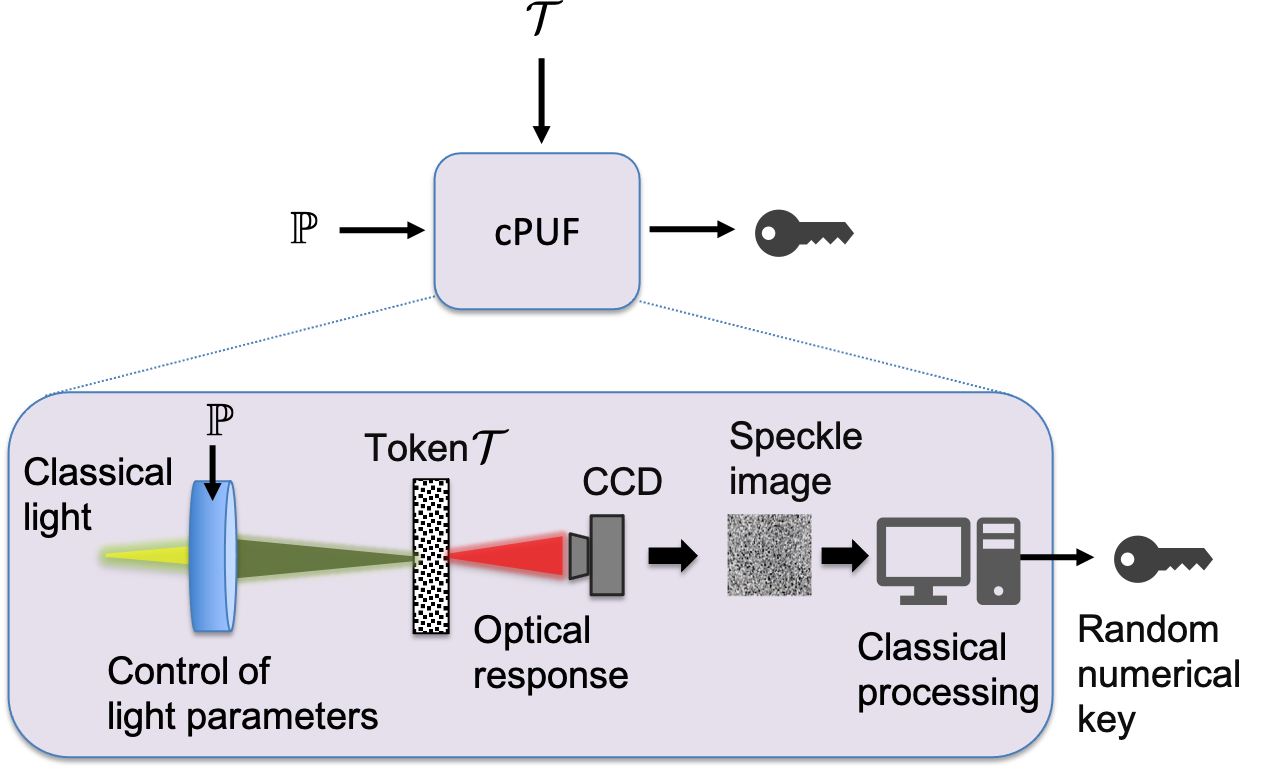}
\caption{Schematic representation of a pseudorandom number generator based on optical PUFs with classical readout (cPUF). Classical light with parameters ${\mathbb P}$ is 
scattered from the disordered token ${\cal T}$. The optical response 
is processed classically to yield a random numerical key. 
Typically, the classical processing also involves error correction so that to ensure 
the stability of the derived key in the presence of inevitable noise and imperfections. The depicted construction can be used for the generation 
of a random binary string, a random integer, or a short (4-digit) PIN, by processing accordingly the speckle. 
}
\label{fig_cpuf}
\end{figure*}

The optical response of the token ${\cal T}$ to classical light  is a random interference pattern (speckle), which depends strongly  
on the internal disorder of ${\cal T}$,  as well as on the parameters of scattered light 
(including  the wavelength, the power, the wavefront, the point and angle of incidence). 
It has been shown that through a judicious classical processing of the speckle, 
one can obtain a random numerical key $K$ \cite{Pappu02,PappuPhD,Mes18,Horstmayer13}, which  passes successfully all of the widely accepted tests for random-sequence  certification (see Fig. \ref{fig_cpuf}). 
This means that for all practical purposes such a key can be 
considered to be close to truly random \cite{Pappu02,PappuPhD,Mes18,Horstmayer13}, and there are no  correlations between different elements or parts of the key, 
as well as between keys  that have been generated from different tokens or from light with different parameters. 
Typically, the processing also involves error-correction so that to ensure stability of the key 
with respect to inevitable innocent noise,  thermal fluctuations, instabilities, etc.
So, the function 
\bea
{\rm cPUF}({\mathcal T}, {\mathbb P}) = K,
\eea
is essentially an {\em optical pseudorandom number generator}, which 
generates the random key $K$ from the token ${\mathcal T}$, when 
seeded with light of parameters ${\mathbb P}$. 
The same token will yield the same random key when 
interrogated by classical light with parameters $ {\mathbb P}$, and thus 
there is no need for storage of $K$, provided one has access to 
both of ${\mathcal T}$ and $ {\mathbb P}$. 
The length of the keys that can be extracted in this way ranges from hundreds to thousands of bits \cite{Pappu02,PappuPhD,Horstmayer13,Mes18}. 

It is important to emphasize once more that 
the classical algorithm used for the extraction of the key from the speckle 
plays a pivotal role in the randomness of the key, and  the wrong choice 
may result in keys which contain correlations and they are far from uniform.

\section{Results} 
\label{secIII}

The EAP we will consider relies on  a judicious combination of the ideas discussed in Refs. \cite{Pappu02,PappuPhD,Mes18,Horstmayer13}. 
At the core of the proposed protocol, there is the optical pseudorandom number generator discussed above.  

\begin{figure*}
\centering\includegraphics[width=13 cm]{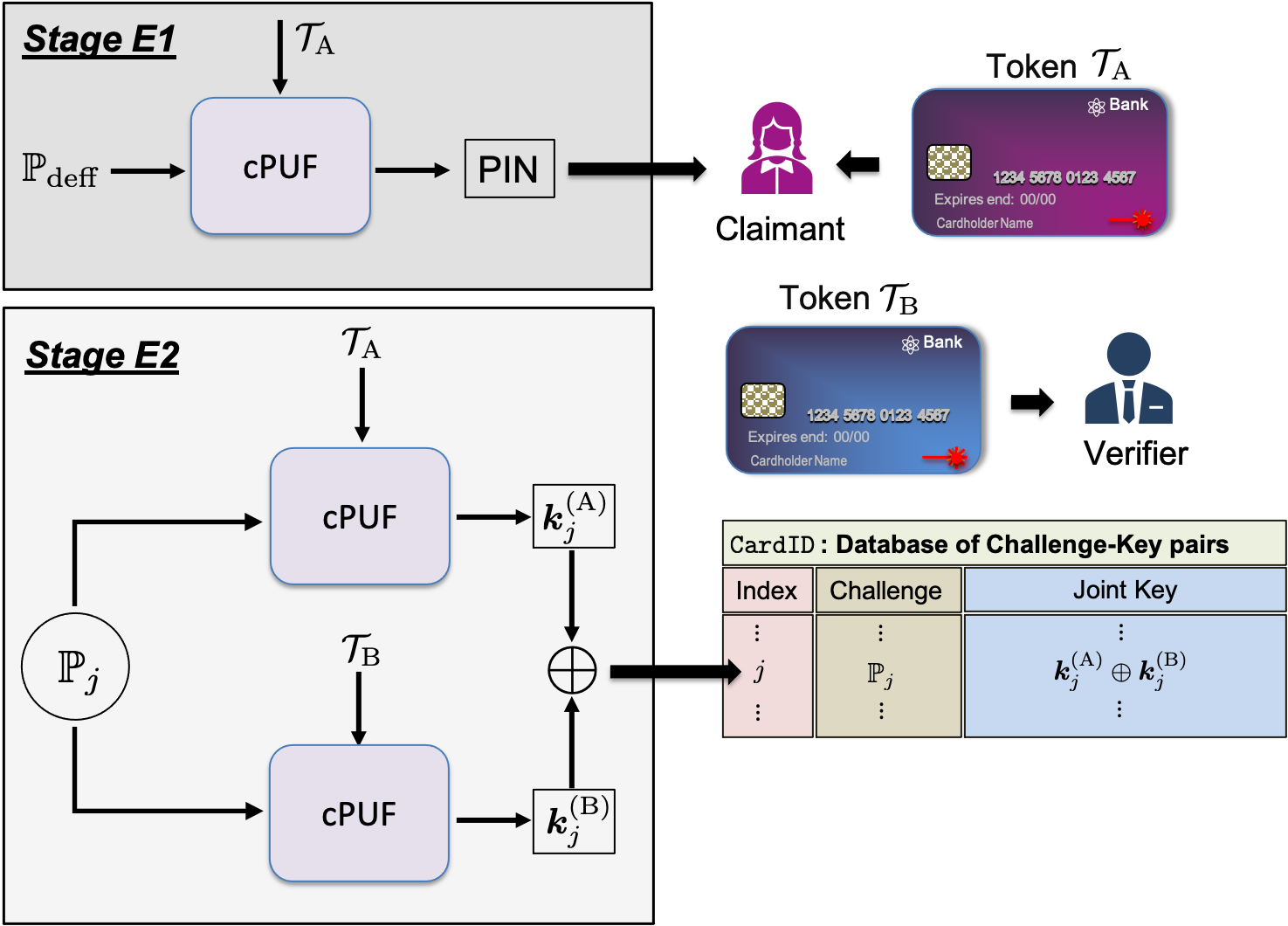}
\caption{Schematic representation of the two stages in the enrollment of a token to the system, for the protocol discussed in 
Sec. \ref{secIII}.  The procedure takes place only once by the manufacturer, in a secure environment.  Stage E1 aims at the generation of a PIN which is correlated with  the token. Stage E2 aims at the generation of a database 
of challenge-response pairs for the particular token. The PIN and the token ${\cal T}_{\rm A}$ are given to the legitimate user, whereas the parameters of the light ${\mathbb P}_{\rm deff}$ used in stage E1  are publicly known. 
}
\label{fig_cd}
\end{figure*}

\subsection{Enrollment of the token to the system}
\label{secIIIA}

We consider a two-stage  enrollment, which  takes place only once, before the token ${\cal T}_{\rm A}$ is given to the legitimate user (Alice). The process in summarized in Fig. \ref{fig_cd}.

{\bf \em Stage E1.---}
The first step aims at the generation of a short (typically four-digit) PIN, 
which is associated with the token. More precisely, the token is interrogated by classical light with a fixed and publicly known set of parameters ${\mathbb P}_{\rm def}$, which may include  wavelength, wavefront, power, position and angle of incidence, etc. This is the default set of parameters for all of the verification set-ups where the user may insert her token. 
To associate the token with a short random PIN we exploit the ideas discussed in Sec. \ref{secII}. 
More precisely, Alice's PIN is obtained by means of the  pseudorandom number generator shown in 
Fig. \ref{fig_cpuf} (with the appropriate classical processing), 
and  for all practical purposes it can be considered stable and close to truly random. 
The PIN is given to Alice together with the associated  token  ${\cal T}_{\rm A}$, and in analogy to conventional smart cards, Alice has to memorize it. By contrast to conventional smart cards, in our case the PIN is not stored on the token or anywhere else. 

{\bf \em Stage E2.---} The second step of the enrollment stage also exploits the ideas discussed in 
Sec. \ref{secII}, and aims at the creation of  a database of challenge-response pairs for the 
token ${\cal T}_{\rm A}$, which will be used by the verifier for the authentication of Alice to the system 
(see below). Let $\{{\mathbb P}_j, j=1,2,\ldots\}$  denote different challenges. Each challenge pertains to the parameters of the classical light to be used in the interrogation of the token, and by contrast to ${\mathbb P}_{\rm def}$ used above, it is not publicly known. These parameters may include wavelength, angle and position of incidence, wavefront, etc.  In order to minimize the 
 secret information that has to be shared between the verifier and the user, we introduce a second token 
 ${\cal T}_{\rm B}$ which is accessible by the verifier only. The two tokens 
 are independent of each other, and thus their 
 responses to the same challenge are also random and independent.  Let ${\bm k}_j^{\rm (A)}$ and 
 ${\bm k}_j^{\rm (B)}$ denote the numerical binary keys generated from tokens  
 ${\cal T}_{\rm A}$ and  
 ${\cal T}_{\rm B}$ respectively,  for challenge ${\mathbb P}_j$. The joint key 
 ${\bm k}_j^{\rm (A)}\oplus {\bm k}_j^{\rm (B)}$ (the symbol $\oplus$ denotes addition modulo 2), together with the corresponding challenge ${\mathbb P}_j$, are stored in the database, whereas the individual keys are never stored in plain text.

\subsection{Verification}
\label{secIIIB}

\begin{figure*}
\centering\includegraphics[scale=0.7]{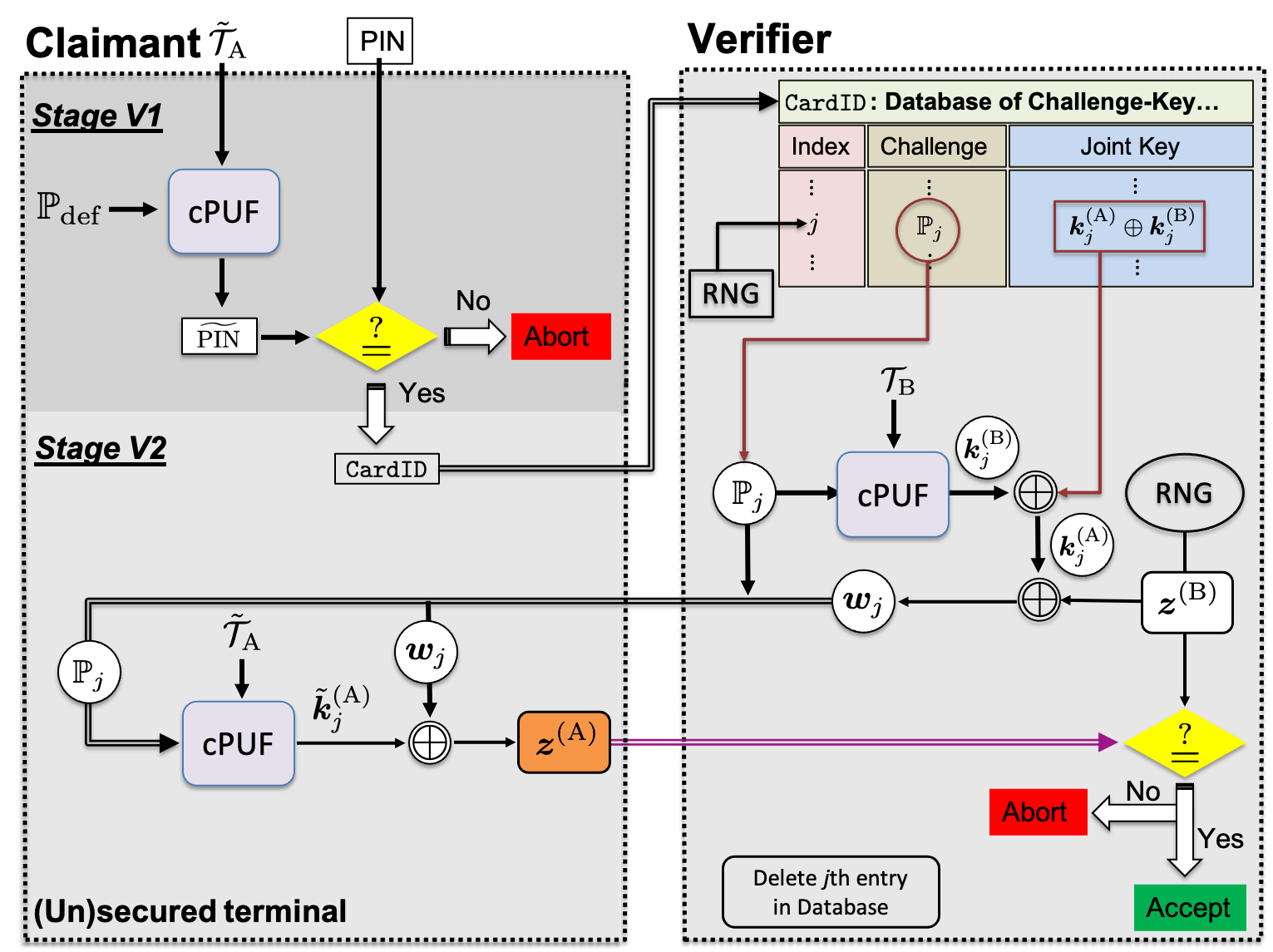}
\caption{Schematic representation of the two stages in the verification of a user (claimant) to the system for the protocol of Sec. \ref{secIII}.  The claimant inserts her token in an (un)secured terminal and types in her PIN. 
Stage V1 takes place locally, and during this stage, the claimant is authenticated to the token. If stage V1 is successful, the terminal contacts the verifier who has access to the database of challenge-key pairs associated with the particular token. Stage V2 aims at the verification of the token, which represents the user to the system. The verifier chooses at random one of the rows in the database, and sends a copy of the associated challenge to the terminal.  In principle, 
a malicious user is allowed to deviate from the depicted procedure, as well as to control various components in the terminal.  The parameters of the light ${\mathbb P}_{\rm deff}$ used in stage V1 are publicly known. Irrespective of the outcome, the data that have been used in the verification are removed from the database at the end of the session. Double-line arrows show classical communication between the verifier and the claimant. 
}
\label{fig_cv}
\end{figure*}

In analogy to entity authentication by means of conventional smart cards, we consider a two-stage verification process, which is summarized in Fig. \ref{fig_cv}. 

{\bf \em Stage V1.---} Initially, the user is authenticated to the token by means of the short PIN, so that to ensure that the token has not been stolen, and it is  possessed by 
the legitimate holder.
By contrast to conventional smart cards, in our case the PIN is not stored on the token, but it can be produced from it easily. As depicted in Fig. \ref{fig_cv}, the claimant  inserts her token to the (un)secured terminal, and types in 
her secret PIN. The token is interrogated by classical light with parameters 
${\mathbb P}_{\rm def}$, and the corresponding response (speckle) is processed classically 
to yield $\widetilde{\rm PIN}$, which is compared to the PIN that the claimant 
has typed in independently. 
The system accepts that the legitimate user is in possession of the token only if  ${\rm PIN} = \widetilde{\rm PIN}$, and it is only in this case that the protocol proceeds to the second stage pertaining to 
the authentication of the token. It is also worth noting that 
stage V1 takes place locally and there is no need for communication with the verifier.

{\bf \em Stage V2.---} Assuming that the verification of the user to the token  has been successful,  the token can represent the claimant 
and the remaining step is its authentication to the system.  

Each authentication session pertains to a different row of the database, which is chosen at random by the verifier (Bob). 
For the sake of concreteness we will consider the $j$th row. Bob uses a classical optical probe with parameters ${\mathbb P}_j$, in order to derive key ${\bm k}_j^{\rm (B)}$ from token 
${\cal T}_{\rm B}$, along the lines discussed in Sec. \ref{secII}. Subsequently, he generates independently a random secret binary message ${\bm z}^{\rm (B)}$ of the same length as ${\bm k}_j^{\rm (B)}$, and sends to the verification set-up the 
ciphertext 
\bea
{\bm w}_j ={\bm z}^{\rm (B)}\oplus ({\bm k}_j^{\rm (A)}\oplus {\bm k}_j^{\rm (B)})\oplus {\bm k}_j^{\rm (B)} = {\bm z}^{\rm (B)}\oplus {\bm k}_j^{\rm (A)}, 
\eea 
together with the associated set of parameters ${\mathbb P}_j$. If the user is honest and has  inserted the right token in the verification set-up, then she should be able to decrypt the message of the verifier. Let us denote by 
 $\widetilde{\cal T}_{\rm A}$ the token that represents the user to the system, which may or 
 may not be Alice's true token. The token $\widetilde{\cal T}_{\rm A}$ is interrogated by classical light 
with parameters ${\mathbb P}_j$, and the resulting speckle is processed classically to yield 
the numerical key $\tilde{\bm k}_j^{\rm (A)}$. All the process takes place locally, and the derived key is XORed  with the received ciphertext ${\bm w}_j$. The resulting message ${\bm z}^{\rm (A)} = \tilde{\bm k}_j^{\rm (A)}\oplus {\bm w}_j$, is sent back to the verifier. 
Bob accepts the token (and thus the claimed identity of the user), if and only if 
 ${\bm z}^{\rm (A)} = {\bm z}^{\rm (B)}$. 
Assuming that innocent noise and imperfections have been taken care by 
the classical processing of the speckle  through which the keys are derived, it is highly unlikely 
for the verifier and the honest user to derive keys which differ from the ones used in the encryption 
of ${\bm z}^{\rm (B)}$ \cite{Pappu02,PappuPhD,Mes18,Horstmayer13}. Hence, any discrepancies between 
${\bm z}^{\rm (A)}$ and ${\bm z}^{\rm (B)}$  can be attributed to a wrong  
token, and/or some type of cheating. 

At the end of the session, the $j$th entry of the database is erased, and it is 
never used again, irrespective of the outcome of the authentication session.

\section{Discussion}
\label{secIV}

We emphasize that in order for Eve to successfully impersonate Alice, she has to pass successfully both of the verification stages discussed in Sec. \ref{secIII}. 

The security of the protocol  stems directly from the 
security of the protocols discussed in Refs. \cite{Pappu02,PappuPhD,Horstmayer13,Mes18}. For the sake of 
completeness it is worth summarizing here the three main cornerstones. The first 
one is the technological hardness of the cloning of the optical 
disordered token. 
This is a common prerequisite for any useful PUF-based 
cryptographic protocol, and there have been some attempts for the quantification of the hardness in the case of optical tokens  \cite{Pappu02,PappuPhD,NikDiaSciRep17,BrussPRA20}. 
The second 
cornerstone  is the  strong sensitivity of the speckle to the internal disorder of the token and  to the parameters 
of the input light, which has been demonstrated in different experimental set-ups   and for various combinations of parameters \cite{Pappu02,PappuPhD,Horstmayer13,Mes18}. The third cornerstone pertains to the 
algorithms used for the conversion of the random speckle to a numerical key. 
These algorithms should be able to convert the random speckle into random numerical key, without introducing any correlations. Various algorithms have been discussed in the literature, and most of them achieve this goal. 

Assuming that these three conditions are satisfied simultaneously, 
then it is highly unlikely for Eve to have both a clone of Alice's token and the associated PIN, in order to 
impersonate her successfully. The unclonability of the token essentially implies that  
 the creation of a nearly perfect clone, which produces the same challenge-response 
pairs as the actual token, is a formidable challenge for current as well as near-future technology. However, given that one cannot be sure about the technology of a potential adversary, it is very important for Alice not to leave her token unattended for a long period of time. The strong dependence of the speckle on the internal disorder of the token as well as on the details of the input light, 
implies that the speckle associated with different tokens and/or different parameters are random and  independent. For judiciously chosen algorithms, 
these properties are transferred over to the derived numerical keys, 
which have been shown  to pass successfully standard randomness tests \cite{Pappu02,PappuPhD,Horstmayer13,Mes18} and thus, 
for all practical purposes, they can be  assumed to be  close to truly random with uniform distribution.

A short 4-digit PIN extracted by truncation of a long  random  numerical key is also close to truly random with 
uniform distribution. It is  worth noting that the PIN verification takes place locally, and  
the PIN never leaves the verification set-up. 
In analogy to conventional smart cards, it is important for the legitimate user to cover the keypad of the set-up while entering her PIN, so that to protect it from an adversary who has installed a camera that monitors the user's actions, without being detected. Under these conditions, the best an adversary can do is to make a guess for the PIN, with the probability of successful guessing being $10^{-4}$.

The keys ${\bm k}_j^{\rm (A)}$ and ${\bm k}_j^{\rm (B)}$ that are used in the token authentication are never stored 
in plain text. Based on the above, they are uniformly distributed independent random strings, because they
pertain to different independent tokens. Even if an adversary has obtained a copy of the database, the 
derivation of the individual keys from ${\bm k}_j^{\rm (A)}\oplus {\bm k}_j^{\rm (B)}$ is impossible by 
virtue of the one-time-pad encryption, unless he has also access to at least  
one of the tokens, either  ${\cal T}_{\rm A}$ or ${\cal T}_B$.  This implies that ${\bm k}_j^{\rm (A)}$ is a secret 
random key, and thus the encryption of the  random message ${\bm z}^{\rm (B)}$  is information theoretically secure.  
An adversary who monitors the communication line that  connects the verification set-up to the verifier,  cannot impersonate successfully the legitimate user, without access to ${\cal T}_{\rm A}$. 
The best she can do is to make a guess for ${\bm z}^{\rm (B)}$  (or equivalently ${\bm k}_j^{\rm (A)}$), and the probability for correct guessing drops exponentially with its length. For instance, when $|{\bm z}^{\rm (B)}| \geq 100$ the probability 
of correct guessing is $2^{-100}\simeq 10^{-30}$. 

In closing, it should be emphasized that the protocol  is not secure if the same row of the database is used in more than one authentication sessions. This is because 
 an adversary who follows the communication between the legitimate  user  and the verifier, can extract ${\bm k}_j^{\rm (A)}$  as follows  ${\bm k}_j^{\rm (A)}={\bm w}_j\oplus {\bm z}^{\rm (A)}$, which allows her to impersonate successfully 
 the legitimate user later on, if the same key is used again (replay attack). This is because ${\bm w}_j$ and ${\bm z}^{\rm (A)}$ are sent in 
 plain text, whereas for the legitimate honest user ${\bm z}^{\rm (A)}={\bm z}^{\rm (B)}$.  In order to prevent such an  attack, and to ensure the security of the protocol for future authentication sessions, it is important for the entire $j$th row to be deleted permanently from the database at the end of the running session, irrespective of the outcome. The number of entries in the database is essentially limited by the number of different keys that can be extracted from a given token, and it is getting smaller after each session. The precise number of entries depends strongly on the type of the optical token, as well as on the parameters of the input light that are exploited
by the protocol. Related studies suggest that 
an optical PUF may support hundreds to thousands different 
challenge-response pairs \cite{Pappu02,PappuPhD,Horstmayer13,Mes18}.  
When the entries in the database are 
exhausted, a new token has to be assigned to the legitimate user. 
Finally, the protocol does not provide protection against 
privileged insiders or superusers, who have access to all of the system's files and resources, including the database of challenge-response pairs 
and the token ${\cal T}_{\rm B}$.  

\section{Conclusions}
\label{secV}

We have discussed remote authentication of entities with 
physical unclonable functions. The protocol  we have presented offers  security against both classical and quantum adversaries, it can be performed with today's technology, and it is  compatible with the infrastructure of standard QKD protocols. 
Only classical information is exchanged between the verifier and the  verification set-up, and thus there are no 
limitations on the distances over which entity authentication can be performed. 
However, the number of different authentication sessions that can be 
performed with the same random token is limited by the amount of randomness that can be extracted from the token, because the 
number of available challenge-response pairs gets smaller after each session, irrespective of the outcome. 
The removal of the used entries from the database is not important for the running authentication session, but it ensures  the security of future  sessions. 

Throughout this work, we have assumed that the verifier is honest. To the best of our knowledge, all of the optical PUF-based EAPs that have been discussed in the literature so far rely on the same assumption. One may relax this assumption, using  quantum PUFs, which however are not compatible with today's technology \cite{qPUF1,qPUF2}.  The development of practical  schemes for remote entity authentication, which offer security even against privileged insiders or superusers, remains an open question. 

Another cryptographic task, which has not been discussed adequately in the literature of PUFs is message authentication \cite{book1}.  
It has been shown recently for a large family of quantum message-authentication codes  \cite{qmac} , that quantum resources do not offer any 
advantage   relative to conventional unconditionally secure message-authentication codes. 
An interesting question therefore is whether optical PUFs can 
offer some advantage, thereby opening up the way to novel 
PUF-based message authentication codes.

\end{document}